\documentclass[aps,prl,preprint]{revtex4-1}
\usepackage{graphicx}
\usepackage{mhchem}

\begin{document}
\title{Adsorption of cationic polyions to a hydrophobic surface in the presence of Hofmeister salts}

\author{Alexandre P. dos Santos}
\email{alexandre.pereira@ufrgs.br}
\affiliation{Departamento de F\'isica, Universidade Federal de Santa Catarina, 88040-900, Florian\'opolis, Santa Catarina, Brazil}

\author{Yan Levin}
\email{levin@if.ufrgs.br}
\affiliation{Instituto de F\'isica, Universidade Federal do Rio Grande do Sul, Caixa Postal 15051, CEP 91501-970, Porto Alegre, RS, Brazil}

\begin{abstract}

We study, using extensive Monte Carlos simulations, the behavior of cationic polyelectrolytes near hydrophobic surfaces in solutions containing Hofmeister salts. The Hofmeister anions are divided into kosmotropes and chaotropes. Near a hydrophobic surface, the chaotropes lose their solvation sheath and become partially adsorbed to the interface, while the kosmotropes remain strongly hydrated and are repelled from the interface.   If the polyelectrolyte solution contains chaotropic anions, a significant adsorption of polyions to the surface is also observed.  On the other hand, the kosmotropic anions have only a small influence on the polyion adsorption.  These findings can have important implications for exploring the antibacterial properties of cationic polyelectrolytes.

\end{abstract}

\maketitle

\section{Introduction}

The study of electrolytes near interfaces is of great importance for science, technology, and medicine.   One  particularly intriguing phenomenon is the repeated 
appearance of the Hofmeister series of ions in various disciplines.  
The first observation that 
many physi\-cochemical processes depend strongly on specific anions present
inside the electrolyte solution dates back over a century ago to the pioneering work of
Hofmeister~\cite{Ho88}. The original lyotropic series was discovered by studying the ability of different ions to precipitate or stabilize protein solutions~\cite{PeGo04,CuLu06,VrJu06}. Following
the pioneering work of Hofmeister, the same series has been observed in many other areas of science, ranging
from electrochemistry, colloidal science~\cite{PeOr10,DoLe11}, surfactant micellization~\cite{JiLi04}, bacterial growth~\cite{LoNi05}, surface and interfacial tensions~\cite{WePu96,JuTo06}, peptide bonds~\cite{HeVi10}, microemulsions~\cite{KaOl95}, etc. Recently, a theory was proposed that allows one to quantitatively predict the effect of Hofmeister ions on the surface~\cite{LeDo09,DoDi10b,DoLe10}  and the interfacial tensions~\cite{DoLe12a,DoLe13a} of electrolyte and acid solutions.
In this paper we want to explore the effect that  Hofmeister ions have on the interaction between cationic polyelectrolytes and a hydrophobic surface.  The study is motivated by the observation that 
{\it amphiphilic} cationic polyelectrolytes (with both hydrophobic and hydrophilic moieties) have been found  
to act as antibiotics. The electrostatic interactions between the negative charges of the membrane and the positive charges of the polyion, as well as the hydrophobic interactions between the phospholipids and the  polyions, appear to be responsible for the antimicrobial behavior~\cite{CaUl04,LuRi05,KiHe12,HoAk12,FaFa12}.
It is then interesting to explore how purely {\it hydrophilic} polyelectrolytes (without hydrophobic parts) behave near a lipid membrane, and if such polyions can also adsorb to a hydrophobic surface in the presence of different electrolytes. A related  property of polyelectrolytes is their ability to stabilize nanoparticles~\cite{GhJi11,PoNa12}.  This also has some significant industrial and medical applications. In the present work, we will focus our attention on the interaction of  cationic polyelectrolytes with soft hydrophobic surfaces.  
To perform this study, we will rely on the Monte Carlos simulations in a conjunction with the 
accurate interaction potentials obtained in the earlier studies of the interfacial tensions of  electrolyte-oil interfaces.

\section{Theory}

The recently developed polarizable anion theory~\cite{Le09,LeDo09,DoDi10b,DoLe10,DoLe12a,BaSt12,DoLe13a} has proven to describe very accurately the experimental data for surface and interfacial tensions, as well as the electrostatic surface potential of different electrolyte solutions~\cite{WePu96,MaTs01,MaYo07}. The important insight of the theory is that the chaotropic and the kosmotropic ions behave differently near a hydrophobic interface~\cite{CoWa85,ZhCr06,CaMa12}. While the kosmotropes maintain their bulk hydration sheath and are repelled from a hydrophobic interface~\cite{LeMe01}, the strongly 
polarizable chaotropes loose their hydration layer and, on approaching the interface, redistribute their electronic charge so that it remains mostly hydrated in the high dielectric
environment~\cite{Le09}. To solvate an ion in water, it is first necessary to create a cavity into which the ion will be inserted.  Creation of a ``hole" in water, costs significant entropic energy, since it perturbs the structure of the water hydrogen bonds and restricts the motion of water molecules.  If, however, the ion moves out of water,  across the interface,  the perturbation to the water
structure vanishes. For small cavities, the cavitational energy scales with the volume of the void~\cite{LuCh99,Ch05}. For strongly hydrated kosmotropic ions, the cavitational energy is too small to compensate the loss of the hydration sheath and for exposing the ionic charge to the low dielectric environment of the lipid membrane.  On the other hand, for weakly hydrated and highly polarizable chaotropes,
the cavitational and the electrostatic energies become comparable~\cite{ToSt13}.  The theory predicts that these ions can become partially adsorbed to the interface.  This prediction was confirmed by the full \textit{ab initio} simulations, which showed that the potential of mean force (PMF) predicted by the polarizable anion theory for 
\ce{I-} anion agreed quantitatively with the results of the \textit{ab initio} simulations~\cite{BaMu11,BaMu13}. 
We first briefly discuss the interaction potentials necessary to study the behavior of 
Hofmeister ions near a hydrophobic interface.  For more detail and the derivations, the interested reader is referred to the original publications.

The dielectric discontinuity across the interface leads to induced charges which repel
ions from the interface~\cite{DoDi10b,Xu12,JaSo12,Xu13,JaSo13}.  This is different from a conductor which leads to an ion-electrode attraction~\cite{PeKu12}.
The approximate charge-image potential at a distance $z$ from the interface was calculated to be~\cite{DoDi10b}
\begin{eqnarray}\label{Uim}
\beta U_{i}(z)=\left\{
\begin{array}{l}
\beta Wa\dfrac{e^{-2 \kappa (z-a)}}{z} \ \text{ for } z \ge  a  \ , \\
\beta W \dfrac{z}{a} \text{ for } 0 \le z< a \ , \\
0 \text{ for } -a \le z <  0 \ ,
\end{array}
\right.
\end{eqnarray}
where $\kappa=\sqrt{8 \pi \lambda_B c_s}$ is the inverse Debye length, $a$
is the ionic radius, $c_s$ is the salt concentration and $\lambda_B=\beta q^2/\varepsilon_w$ is the Bjerrum length, which for water at room temperature is $7.2$~\AA,  and $q$ is the proton charge. The potential at contact, $W$, is given by~\cite{DoLe13b},
\begin{equation}\label{image_potential_i}
\beta W=\frac{\lambda_B}{2} \int_0^\infty dk \ \frac{k\ f_1(k)}{p\ f_2(k)}  \ ,
\end{equation}
where 
\begin{eqnarray*}
f_1(k)=p \cosh{(k a)}-k \sinh{(k a)}+\nonumber \\
\frac{\epsilon_o}{\epsilon_w}p\sinh{(k a)}-\frac{\epsilon_o}{\epsilon_w}k \cosh{(k a)} \ ,
\end{eqnarray*}
\begin{eqnarray*}
f_2(k)=p \cosh{(k a)}+k \sinh{(k a)} +\nonumber \\
\frac{\epsilon_o}{\epsilon_w}p\sinh{(k a)}+\frac{\epsilon_o}{\epsilon_w}k \cosh{(k a)} \ ,
\end{eqnarray*}
and $p=\sqrt{k^2+\kappa^2}$. The dielectric constants of water and oil  are $\varepsilon_w=80$ and $\varepsilon_o=2$, respectively. This potential
accounts approximately for both the polarization of the ionic atmosphere around each ion and for the induced charge at the dielectric interface.  

In simulations, of course, it is possible to use directly the image charges of
ions and monomers~\cite{LuLi11,DoBa11,Xu13} to account for the induced surface 
charge.  However, as we will see
later, for monovalent ions the approximate potential Eq.~\eqref{Uim} provides us with a major speed up of simulations, without any
significant sacrifice of accuracy.

The hydrophobic potential is proportional to the ionic volume exposed to the aqueous medium and is given by
\begin{eqnarray}\label{cavpot}
\beta U_c(z)=\left\{
\begin{array}{l}
 \nu a^3 \text{ for } z \ge  a  \ , \\
 \frac{1}{4} \nu a^3  \left(\frac{z}{a}+1\right)^2 \left(2-\frac{z}{a}\right)
\text{ for } -a<z<a \ ,
\end{array}
\right.
\end{eqnarray}
where $\nu=0.3/$\AA$^3$ is obtained using SPC/E water simulations~\cite{RaTr05}.

The electrostatic self energy of a chaotropic ion at a distance $z$ from the
interface is~\cite{LeDo09}
\begin{eqnarray}\label{Upol}
\beta U_{p}(z)=\left\{
\begin{array}{l}
\frac{\lambda_B}{2 a} \text{ for } z\ge a \ , \\
\frac{\lambda_B}{2 a}\left[\frac{\pi x^2}{\theta(z)}+\frac{\pi
[1-x]^2 \epsilon_w}{[\pi-\theta(z)]\epsilon_o}\right] + \\
g \left[x-\frac{1-cos[\theta(z)]}{2} \right]^2 \text{ for } -a<z<a  \ ,
\end{array}
\right.
\end{eqnarray}
where $\theta(z)=\arccos[-z/a]$, $g=(1-\alpha)/\alpha$, and $x$ is the fraction of the
ionic charge that remains hydrated. The relative polarizability is defined as $\alpha=\gamma/a^3$, where $\gamma$ is the ionic polarizability. Minimizing Eq.~\eqref{Upol}, we obtain the fraction of the ionic charge that remains hydrated,
\begin{equation}
x(z)=\dfrac{\dfrac{\lambda_B \pi \epsilon_w}{a \epsilon_o
\left[\pi-\theta(z)\right]}+g [1-cos[\theta(z)]]}{\dfrac{\lambda_B
\pi}{a \theta(z)} + \dfrac{\lambda_B \pi \epsilon_w}{a \epsilon_o [\pi-\theta(z)]} +2 g} \ .
\end{equation}

Ions near a hydrophobic surface also experience dispersion interactions which
are modeled by the potential~\cite{DoLe12a}
\begin{eqnarray}
\label{edis}
\beta U_{d}(z)=\left\{
\begin{array}{l}
 0 \text{ for } z \ge  a  \ , \\
 A_{eff} \alpha [1 - \\
\dfrac{(z/a + 1)^2(2 - z/a)}{4} ] \text{ for } -a<z<a \ ,
\end{array}
\right.
\end{eqnarray}
where $A_{eff}=-4.4$ is the effective Hamaker constant~\cite{DoLe12a}.

The chaotrope-interface interaction potential is the sum of all these contributions, $U_i(z)+U_c(z)+U_p(z)+U_d(z)$. Since the kosmotropes do not penetrate the surface, their
interaction potential consists only of the  ion-image contribution, $U_i(z)$, and the
hardcore repulsion (at one hydrated radius) from the interface. 

\section{Monte Carlo Simulations}

Our system consists of cations and anions, derived from a
Hofmeister salt, and of cationic polyions with the dissociated counterions.   
All the particles are confined inside a box of sides $L_x=L_y=182$~\AA\ and $L_z=200$~\AA, centered at the origin of the coordinate system. A soft hydrophobic surface (water-oil interface) is located at $z=0$. The electrolyte and polyelectrolyte are restricted to the region $0<z<L_z/2$. Only chaotropic ions can penetrate into the region $-r_a<z<L_z/2$, where $r_a$ is the anionic radius. The aqueous medium, $z>0$, is considered to be uniform with the 
dielectric constant $\varepsilon_w$. The uncharged lipid
membrane, modeled as oil, occupies the region $z<0$, and has the dielectric constant $\varepsilon_o$. In the spirit of the primitive model, all ions are treated as hard spheres. 
Sodium cation has radius $r_{Na^+}$ and charge $q$,  anions have radius $r_a$ and charge $-q$, while the polyion counterions have charge $-q$ and radius of $\ce{Cl-}$. 
The polyions have $10$ monomers each.    Each monomer is a sphere of radius 
$r_{Na^+}$ and  carries a positive charge $q$.
The monomers interact between themselves with a 
repulsion WCA potential~\cite{StKr95},
\begin{eqnarray}\label{lj}
\beta U_{LJ}(r)=\left\{
\begin{array}{l}
4\epsilon[ (\sigma/r)^{12}-(\sigma/r)^6 -(\sigma/r_c)^{12}+\\
(\sigma/r_c)^6] \text{ for } r < r_c  \ , \\
0 \text{ for } r\ge r_c \ ,
\end{array}
\right.
\end{eqnarray}
where $r$ is the distance between the monomers, $r_c=2^{1/6}\sigma$, $\sigma=2~r_{\ce{Na+}}$, and $\epsilon=0.833$. The  adjacent monomers interact by the FENE attractive 
potential~\cite{StKr95},
\begin{equation}\label{fene}
\beta U_{FE}(r)=-0.5 k_s R_0^2 \ln{(1-r^2/R_0^2)} \ ,
\end{equation}
where $k_s=7 \epsilon/\sigma^2$ is the spring constant and $R_0=2\sigma$ is the maximum extent. We use the same parameters as Stevens and Kremer~\cite{StKr95} for polyelectrolytes.

\begin{figure}[t]
\begin{center}
\includegraphics[width=10cm]{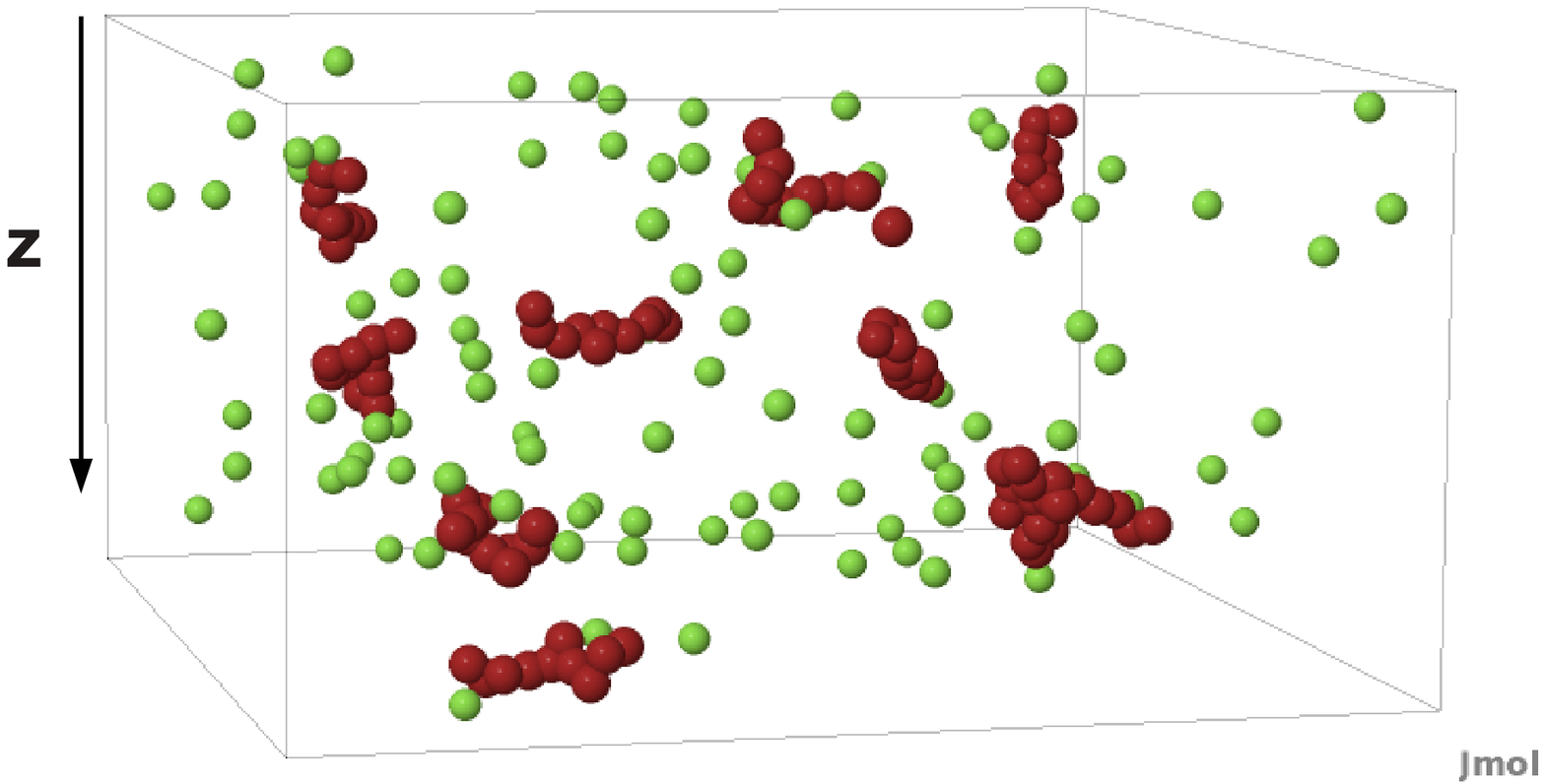}\hspace{0.2cm}\vspace{0.2cm}
\includegraphics[width=10cm]{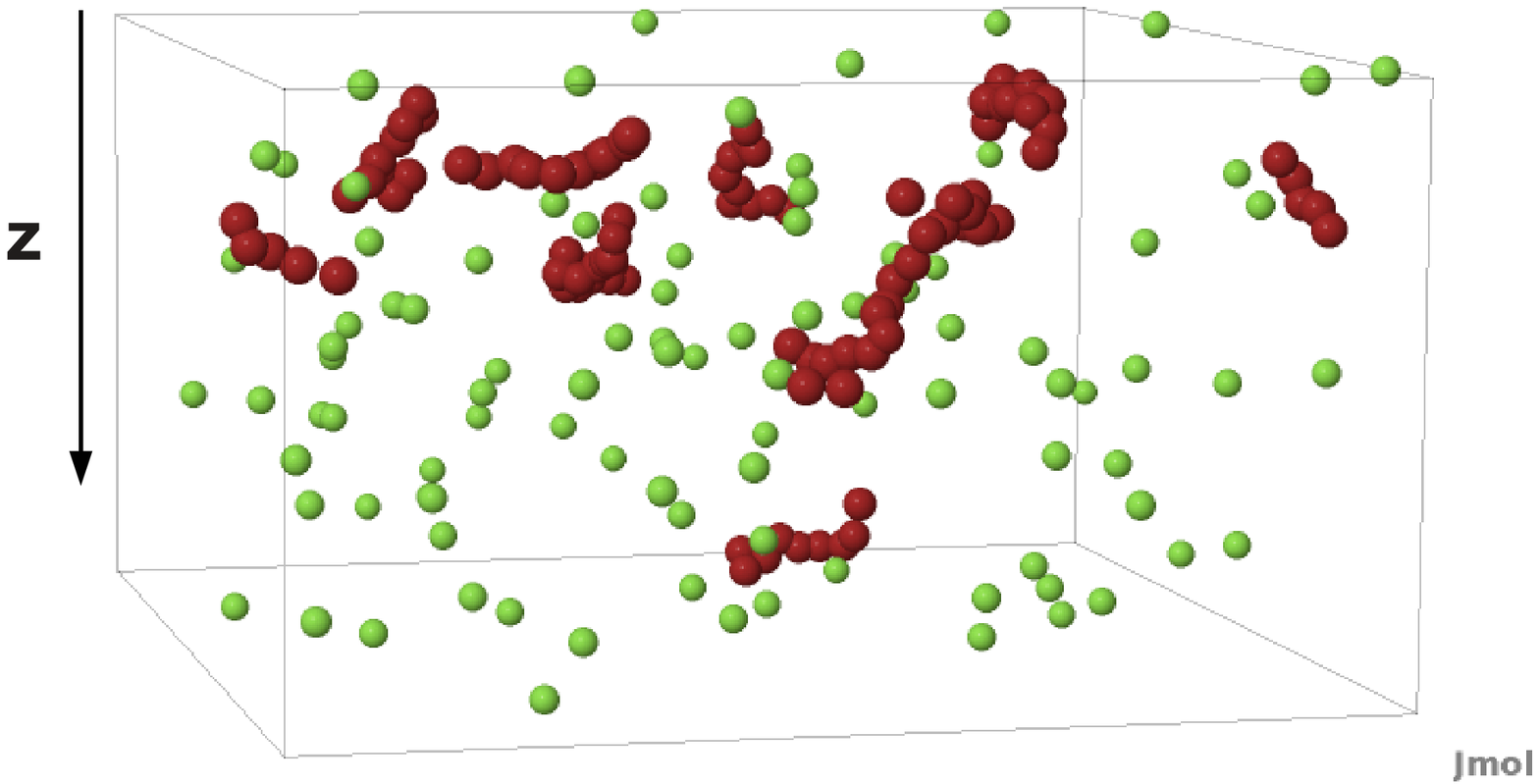}\hspace{0.2cm}\vspace{0.2cm}
\end{center}
\caption{Snapshots of two equilibrium configurations for salts \ce{NaF} (top panel) and \ce{NaI} (bottom panel), both at concentration $50$~mM. The concentration of polyions is $5$~mM. Only anions (lighter spheres) and polyions (darker spheres) are shown. The oil-water interface is at the top of the simulation box.}
\label{fig1}
\end{figure}
All the ions and the monomers interact between themselves with the Coulomb potential. Since in the simulations we use a slab geometry, a correction of Yeh and Berkowitz~\cite{YeBe99} to the standard 3d Ewald summation method is applied. The kosmotropic ions, \ce{Na+}, \ce{Cl-}, \ce{F-},  and the polyion monomers interact with the interface through the effective potential $U_i(z)$, while the chaotropic anions, $\ce{Br-}$ and $\ce{I-}$, interact with the interface through the effective potential $U_i(z)+U_c(z)+U_p(z)+U_d(z)$. In the simulations the dielectric jump at $z=0$ is taken into account through the potential $U_i(z)$, without using explicitly the image charges. We perform regular displacements for all the microions, and reptation and rotation moves for the polyions. The equilibration is achieved after $10^6$ MC steps.   Subsequently, $10$ movements per particle are used to obtain the uncorrelated configurations, and the averages are calculated using $10^4$ equilibrated uncorrelated states.  Fig.~\ref{
fig1} shows the characteristic 
equilibrium configurations for the salts \ce{NaF} and \ce{NaI}.  We see that  \ce{NaI}
results in an enhanced adsorption of the polyions to the interface.

\section{Test of the effective potential}

\begin{figure}[t]
\begin{center}
\includegraphics[width=8cm]{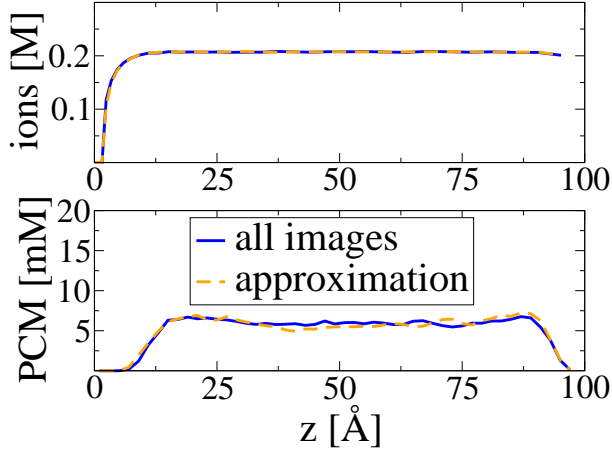}
\end{center}
\caption{Comparison between the density profiles obtained from the ``exact" simulation of Ref.~\cite{DoLe13b} and a simulation using the effective potential $U_{i}(z)$. In the upper panel we show the ionic profiles of the electrolyte \ce{NaCl}, at concentration $0.2$~M. The density profiles for \ce{Na+} and \ce{Cl-} are indistinguishable. In the lower panel we plot the PCM~(polyelectrolyte center of mass) profile for $5$~mM polyelectrolyte and \ce{NaCl} at concentration $0.2$~M.  In both cases the agreement between the ``exact" and the ``approximate" simulations is very good.}
\label{fig2}
\end{figure}

One of the difficulties in 
performing simulations with the dielectric interfaces is the 
need to account for the induced surface charge,  see Ref.~\cite{DoLe13b} for details. 
This makes the simulations
very slow.  For strongly polarizable chaotropic anions there is an additional difficulty since it is
very hard to calculate exactly the electrostatic potential of an ion as it penetrates
the dielectric interface.  To avoid these problems, in the previous section we have used the
effective charge-image interaction potential. This potential was derived by considering 
the solution of the linearized Poisson-Boltzmann equation.  To test the reliability of this 
potential, we first 
study a  polyelectrolyte solution with a kosmotropic salt, \ce{NaCl}.  For kosmotropic
ions  -- which never cross the interface -- it is possible to  account exactly for the induced charges 
by including for each ion or monomer a corresponding image charge. 
We can then compare 
the results of these ``exact" simulations with the ``approximate" simulations, in which the images are not included, but instead each ion and monomer interacts with the surface through an 
effective potential $U_i(z)$, with the inverse Debye length $\kappa=\sqrt{8 \pi \lambda_B c_s + 4 \pi \lambda_B N_m c_p}$, where $N_m=10$ is the number of monomers in a polyion, and $c_p$ is the polyelectrolyte concentration. Fig.~\ref{fig2} shows that the simulations with
the effective potential $U_i(z)$ produce the ionic
and polyelectrolyte density profiles practically indistinguishable from the exact all image simulations.  
With the simulation method thus validated for kosmotropic ions, we can now proceed to study general polyelectrolyte
solutions containing a full spectrum of Hofmeister ions.

\begin{figure}[h]
\begin{center}
\includegraphics[width=8cm]{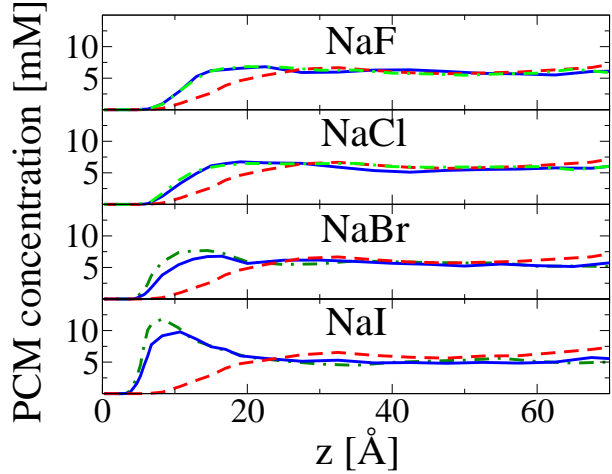}
\end{center}
\caption{Polyelectrolyte center of mass~(PCM) density profile as a function of the distance $z$ from the interface, for various electrolytes and concentrations. Polyelectrolyte concentration is $5$~mM. Dashed (red), solid (blue), and dashed-dotted (green) lines correspond to solutions containing salts at concentrations $0$, $0.2$, and $0.3$~M, respectively.}
\label{fig3}
\end{figure}

\section{Results}

In simulations, 
all the ionic sizes and the polarizabilities are the same as in the previous works on surface
and interfacial tensions of electrolyte solutions. The hydrated radii of the 
kosmotropic ions, $\ce{F-}$ and $\ce{Cl-}$, are $r_a=3.54$~\AA\ and $r_a=2$~\AA, respectively~\cite{Ni59}. The chaotropic ions, such as bromide and iodide, loose their hydration sheath
and as the result of their high polarizability can become partially adsorbed to  hydrophobic surface.   The bare radius of $\ce{Br-}$ is $r_a=2.05$~\AA\ and its  polarizability is $\gamma=5.07$~\AA$^3$~\cite{LaPi39,PyPi92}.  The bare radius of $\ce{I-}$ is $r_a=2.26$~\AA\ and its polarizability is  $\gamma=7.4$~\AA$^3$~\cite{LaPi39,PyPi92}. The 
hydrated radius~\cite{DiDo12} of  \ce{Na+} is $r_{Na^+}=1.8$~\AA.

\begin{figure}[b]
\begin{center}
\includegraphics[width=8cm]{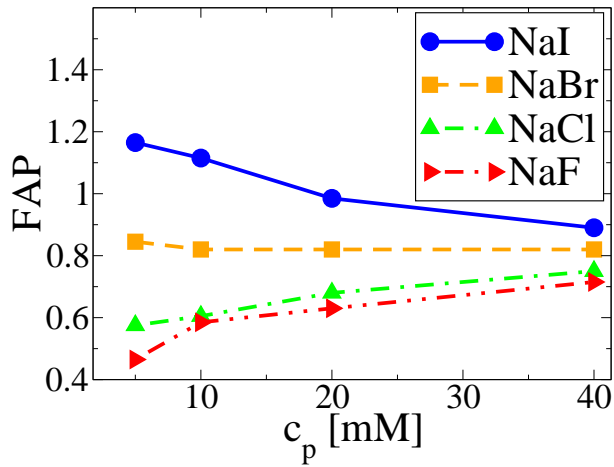}
\end{center}
\caption{Fraction of adsorbed polyions~(FAP) as a function of polyelectrolyte concentration, at salt concentration of $0.2$~M.}
\label{fig4}
\end{figure}
The simulations show that addition of kosmotropic salts such as \ce{NaF} and \ce{NaCl} does
not have a significant influence on the interfacial adsorption of cationic polyions, see Fig.~\ref{fig3}. The only effect of the kosmotropic salts is to increase the electrostatic screening of the monomer-image interaction, allowing the polyions to come
closer to the interface. 
The situation is very different if solution contains chaotropic anions.  In this case, there is a significant adsorption of polyions to a hydrophobic interface, Fig.~\ref{fig3}. The adsorption is stronger for 
larger chaotropic ions, such as iodide.  

In Fig.~\ref{fig4}, we plot the  fraction of adsorbed polyions~(FAP) as a function of the polyelectrolyte concentration. FAP is defined as the concentration of adsorbed polyions ( within the distance  $z=20$~\AA\ of the interface) divided by the average concentration of the polyions inside the solution. We see that if solution contains chaotropic anions, at low concentrations of polyelectrolyte, a significant number of polyions is adsorbed to the interface. 
The behavior is different for electrolytes containing chaotropic and kosmotropic anions. 
For chaotropes, the FAP decreases with the polyelectrolyte concentration. This happens because at
higher concentration of polyelectrolyte, there are will be more dissociated 
counterions, which leads to an increased screening
of the electrostatic attraction between the adsorbed chaotropic anions and 
the polyions of the solution. 
On the other hand, when the polyelectrolyte solution contains only 
kosmotropic ions (which are repelled from the interface),
increased polyelectrolyte concentration leads to stronger screening of charge-image repulsion, 
allowing the polyions to
approach closer to the hydrophobic surface --- leading to an increase of FAP as a function of the polyelectrolyte concentration.  

\section{Conclusions}

We have studied the interactions of cationic 
polyelectrolytes with a hydrophobic surface in the presence of Hofmeister salts. 
Ionic interactions with the interface were modeled using the  
effective potentials derived in the earlier 
studies of the  interfacial tension of electrolyte-oil interfaces.  
For salts containing kosmotropic anions,
the simulation methodology was verified against the explicit all-image simulations.
This opens a possibility of using the effective  potentials to 
significantly speeds up the simulations for systems containing monovalent ions.  
The agreement between the ``approximate" and the ``exact" simulations also suggests that the effective 
potentials should be sufficient to study polyelectrolyte solutions containing chaotropic anions.  
In this case, the direct implementation of the ``exact" simulations is further complicated
by the difficulty of accounting for the electrostatic boundary conditions 
when strongly polarizable chaotropic ions penetrate the dielectric interface.  
The method presented in this paper allows us to overcome this problem by 
using the effective potentials.  At this point, however, our results for chaotropic ions are a prediction based on
the effective potentials developed in the earlier studies of the oil water interface~\cite{DoLe12a} and should be
verified by more detailed atomistic simulations~\cite{VaPl12}.  Unfortunately, the state of the art force fields for atomistic simulations
lead to too much adsorption of chaotropic ions, resulting in surface tensions which deviate strongly from the experimental
measurements~\cite{ToSt13,DoLe13a}.  On the other hand, the effective potentials used in the present work result in a smaller adsorption and
produce surface and interfacial tensions in excellent agreement with the experimental measurements~\cite{LeDo09,DoDi10b,DoLe12a}.  This makes it difficult for us to compare our methodology with the classical atomistic simulations.

An important conclusion of the present work is that 
the chaotropic anions can significantly increase the adsorption of the 
cationic hydrophilic polyions to an uncharged 
lipid membrane.  The simulations presented here allow us to make quantitative estimates on the amount of adsorption. Since the antibacterial effect of a polylectrolyte is related to the polyion adsorption, this can prove to be important for the development of antibacterial applications involving hydrophilic cationic polyelectrolytes --- the amphiphilic polyions used for this purpose often show high toxicity. However, since the bacterial membranes are partially 
composed of negatively charged phospholipids, 
it will also be important to explore the role of the surface charge 
on the polyion adsorption.  

\bibliography{ref.bib}

\end{document}